
\magnification=\magstep1
\vsize=23truecm
\hsize=15.5truecm
\hoffset=.2truecm
\voffset=.8truecm
\parskip=.2truecm

\font\ti=cmbx10 scaled\magstep1
\font\eightrm=cmr8
\font\ninerm=cmr9
\def\br{\hfill\break\noindent}
\def\ra{\rightarrow}
\def\u{\underline}

\def \G11{\Gamma_{11}}
\def\m {\cal M}
\def\n {\cal N}
\def\p {\cal P}
\def \cl {{\rm Cliff }\bigl( SO(10)\bigr)}
\def \A {\cal A}

\def \s {SO(10)}
\def \a {\alpha }
\def \b {\beta }
\def \ah {\hat{\alpha }}
\def \bh {\hat {\beta }}
\def \ot {\otimes}
\def \g5{\gamma_5}
\def \ra{\rightarrow}
\def \l4{\Bigl( {\rm Tr}( KK^*)^2-({\rm Tr}KK^*)^2\Bigr)}
\def \bp{\oplus }
\def \k2{{\rm Tr}KK^*}
\def \slash#1{/\kern -6pt#1}
\def \di{\slash{\partial}}

\pageno=0

%
%
\baselineskip=.5truecm
\footline={\hfill}
{\hfill ZU-TH- 10/1993}
\vskip.1truecm
{\hfill ETH/TH/93-12 }
\vskip.2truecm
{\hfill 12 March 1992}
\vskip2.1truecm
\centerline{\ti SO(10) Unification in Non-Commutative Geometry }
\vskip1.2truecm
\centerline{  A. H. Chamseddine$^{1}$ \footnote*
{\ninerm Supported in part by the Swiss National Foundation (SNF)}
 and J. Fr\"ohlich$^{2}$ }
\vskip.8truecm
\centerline{$^{1}$ Theoretische Physik, Universit\"at Z\"urich, CH
8001 Z\"urich Switzerland}
\centerline{$^{2}$ Theoretische Physik, ETH, CH 8093 Z\"urich Switzerland}
\vskip1.2truecm

\centerline{\bf Abstract}
We construct an $SO(10)$ grand unified theory in the formulation
of non-com-\break mutative geometry.
The geometry of space-time is that of a product of a continuos
four dimensional
manifold times a discrete set of points. The properties of the
fermionic sector fix almost uniquely the Higgs structure.
The simplest model corresponds
to the case where the  discrete set consists of three points  and
the Higgs fields are ${\u {16}_s}\times \overline{\u {16}}_s$ and
${\u {16}_s}\times {\u {16}_s} $. The requirement
that the scalar potential for all the Higgs fields  not vanish
imposes strong restrictions on the vacuum expectation values of the
Higgs fields and thus the fermion masses.
We  show that it is possible to remove these constraints
by extending the number of discrete points to six and adding a singlet
fermion and a ${\u {16}_s}$ Higgs field. Both models are studied
in detail.

\vskip.5truecm

\noindent

\vfill

\eject
\baselineskip=.6truecm
\footline={\hss\eightrm\folio\hss}

\centerline{}
\vskip1.1truecm
{\bf\noindent 1. Introduction}
\vskip.2truecm
\noindent
Grand unified theories provide an attractive mechanism to unify
the weak, strong and electromagnetic interactions and put
order into the representations of quarks and leptons. At present,
the simplest models are based on  $SU(5)$ [1] and
$SO(10)$ gauge theories [2]. The second class of models has the
advantage of
including all the fermions (plus a right handed neutrino) in
one representation. This advantage does not translate itself into
a more predictive theory, because there are many possibilities to
break $SO(10)$ down to $SU(3)\times U(1)_{\rm em}$ requiring many
different and often complicated Higgs representations [3]. What is
clearly needed in grand unified theories is a principle to put  order into
the Higgs sector. During the last few years, much effort has
been directed
towards this problem by studying unified theories as  low-energy
limits of the heterotic string . Although this is an attractive strategy,
it has proven to be a difficult one, due to the fact that one must search
for  good models among the very large number of string vacua.
We shall follow, instead, a different strategy.

It has been shown by Connes [4-5] and Connes and Lott [6-7]
that the ideas of non-commutative geometry
can be applied to, among other things,  model building in
particle physics. In
particular, the Dirac operator, defined on the one-particle
Hilbert space of quarks
and leptons, is used to construct the standard $SU(3)\times SU(2)
\times U(1)$ model with the Higgs field unified with the gauge fields.
The space-time used in this constrution is a product of a
Euclidean four-dimensional manifold by a discrete two-point space.
If, in  coming years, an elementary Higgs field is observed
experimentally, one can turn the argument around and view it as an indication
that space-time has the product structure proposed by Connes. One
expects that this beautiful and highly symmetric construction
would yield some predictions, in particular constraints among the coupling
constants and particle masses, and indeed it does under certain
circumstances [8]. (See also [9] for an alternative realization
of Connes program). However, such relations can only be taken seriously
once quantization is understood, or if one can stabilize the
radiative corrections by, for example, supersymmetrising the theory.
In a recent paper [10] it has been shown that, by a simple modification of
the construction of Connes, it is possible to obtain unified models
such as the $SU(5)$ and left-right $SU(2)_{\rm L}\times
SU(2)_{\rm R}\times U(1)_{{\rm B}-{\rm L}}$ theories. Other models such as
the flipped $SU(5)\times U(1)$ model are also within reach of these
constructions.
The interesting case of $SO(10)$ was not treated, because it was
not clear how to proceed in view of the fact that a realistic
$SO(10)$ model requires  complicated Higgs representations.
Meanwhile, it has turned out that the solution is fairly simple,
and the construction
of a realistic $SO(10)$ model will be the main concern of this paper.
All the tools
that will be used here are explained in references [10], and a self contained
summary can be found in section 2 of the second item in reference [10]
(The results
contained there will be freely used in this paper.)

The plan of this paper is as follows. In section 2, we construct
the Dirac operator associated with an $\s $ gauge theory and show
that the simplest model corresponds to a discrete space of three points.
In section 3, the symmetry breaking chain is described in detail
and the vevs of the Higgs fields are given. In section 4, the potential
is analyzed, and it is shown that a potential  survives after
eliminating the auxiliary fields only if the vevs of the Higgs fields
satisfy certain constraints. In section 5, we show that it is possible
to relax the constraints, provided that the number of discrete
points is taken to be six and certain symmetries are imposed.

{\bf \noindent 2. The SO(10) framework }
\vskip .2truecm
\noindent
The starting point in  Connes' construction [4-8] is the specification
of the fermionic sector and the Dirac operator on the space of
spinors. In the $SO(10)$-model [2], the fermions  neatly fit in the
$\underline {16}_{\rm s}$ spinor representation, repeated three times.
A single fermionic family is described by the field $\psi_{\a \ah}$,
where $\a $ is an $SO(1,3)$ Lorentz spinor index with four
components and $\ah $ is an
$SO(10)$ spinor index with thirty two components.
It satisfies both space-time and $SO(10)$
chirality conditions:
$$
\eqalign{
(\g5 )_{\a}^{\b}\psi_{\b\ah}&=\psi_{\a\ah} \cr
(\G11 )_{\ah}^{\bh}\psi_{\a\bh}&=\psi_{\a\ah}.\cr}\eqno(2.1)
$$
where $\g5 =i\gamma_0\gamma_1\gamma_2\gamma_3 $,
$\G11 =-i\Gamma_0\Gamma_1 \cdots \Gamma_9 $, and for later
convenience we have denoted $\Gamma_{10}$ by $\Gamma_0 $.
This reduces the independent spinor components to two for
the space-time indices, and to sixteen for the $\s$ indices.
The general fermionic action is given by
$$
\overline {\psi_{\a\ah}^p}\bigl( \di +A^{IJ}\Gamma_{IJ}\bigr)_{\a\ah}
^{\b\bh}\psi_{\b\bh}^p \ +\psi_{\a\ah}^{Tp}C^{\a\b}H_{\ah\bh}^{pq}
\psi_{\b\bh}^q \eqno(2.2)
$$
where $C$ is the charge conjugation matrix,  $p, q=1,2,3$ are
family indices, and $H$ is some appropriate combination of
Higgs fields breaking the subgroup $SU(2)\times U(1)$
of $SO(10)$ at low energies. An exception of a Higgs
field that  breaks the symmetry at high energies and yet couples to
fermions is the one
that gives a Majorana mass to the right handed neutrinos [11]. The
other Higgs fields needed to break the $SO(10)$ symmetry at
high energies should not couple to the fermions so as not to
give the quarks and leptons super heavy masses.

{}From the form of eq.(2.2) we deduce that the gauge and Higgs fields
are valued in the Clifford algebra of $SO(10)$, projected with
the chirality operator acting on the right and the left of the fields.
Since we know that in the non-commutative construction the Higgs

fields are obtained by having more than one copy of
Minkowski space, we need to choose a discrete space
containing at least three points. On two of the copies, the
associated spinors are taken to be identical, and the Higgs fields
in this direction will not couple to the fermions as these have
the same chirality. On the third copy the fermions are taken to be
the conjugate spinors, as can be deduced from the second term of
eq (2.2). Thus, between copies one and two, we must impose a
permutation symmetry, while between copies one and three we must
require some form of conjugation symmetry. If we insist that the
fermionic sector exhibit a $Z_2$-symmetry then four copies of
Minkowski space are necessary, with the third and fourth copies
identified, too.
 This option will be pursued
in the last section. Since both $SO(1,3)$ and
$\s $ have conjugation matrices, we take the conjugate spinor
to be given by
$$
\psi^c \equiv BC \overline {\psi }^T  \eqno(2.3)
$$
where $B$ is the $\s $ conjugation matrix satisfying
$B^{-1}\Gamma_I B =-\Gamma_I^T $. Thus the spinor for the
system is given by
$$
\Psi =\pmatrix{ \psi \cr \psi \cr \psi^c \cr } \eqno(2.4)
$$
The chirality conditions on the spinor $\Psi $ are given by
$$\eqalign{
\g5 \ot {\rm diag }(1,1,-1) \Psi &=\Psi \cr
\g5 \ot \G11 \Psi &=\Psi  \cr}\eqno(2.5)
$$
Before proceeding, it is useful  to address the
problem of  neutrino masses. The right handed neutrino
must acquire a large mass. This is usually done by coupling the
fermions to a $\underline{126} $ or to a $\underline{{16}_{\rm s}}$
Higgs field with appropriate vacuum expectation values (vev's)
giving a mass to the right handed neutrino but not to the remaining
fermions. The ${\u {126}}$ appears already with the Higgs fields
that give masses to the fermions.  The $\underline {16}_{\rm s}$ can only
be obtained by extending the fermionic space with a singlet
spinor. This implies that the number of copies of Minkowski space
must be increased by one or two, depending on whether the
$Z_2$ symmetry is required or not.
In this case, two of the neutral fermions will become superheavy,
while the third would  remain massless.
The fermionic space is then chosen to be
$$
 \pmatrix {\psi \cr \psi \cr \psi^c \cr \psi^c
 \cr \lambda \cr \lambda^c \cr} \eqno(2.6)
$$
where the number of copies associated with conjugate spinors
is doubled.
We shall first consider a spinor space corresponding to eq. (2.4)
and treat the more complicated case corresponding to eq. (2.6) in
the last section.

We are now ready to specify a triple $ ({\A} ,h,D) $ defining a
 non-commutative geometry, where $h$ is the Hilbert space of the
spinors $\Psi $, ${\A} $ is an involutive algebra of operators on $h$,
and $D$ is an unbounded, self-adjoint operator on $h$ [4-5]. Let $X$ be
a compact Riemannian four-dimensional spin-manifold, ${\cal A}_1 $
the algebra of functions on $X$ and $(h_1,D_1,\Gamma_1 )$ the
Dirac K cycle, with $h_1\equiv L^2 (X, \sqrt g d^4x )$, on ${\A}_1
$, and $\Gamma_1 $ is a $Z_2$ grading.
We choose ${\A} $ to be given by
$$
{\A}_2 =P_+ {\cl }P_+  \eqno(2.7)
$$
where $P_{\pm}={1\over 2} (1\pm \Gamma_{11}) $, and set
$$
{\A} ={\A}_1 \ot {\A}_2
$$
We define $\Omega^* ({\A })={\bp}_{n=0}^{\infty} \Omega^n ({\A }) $
to be the universal differential algebra over ${\A}$, with
$\Omega^0 ({\A })={\A} $, and
$$
\Omega^n ({\A})=\{ \sum_i a_0^i da_1^i\ldots da_n^i \ : a_j^i \epsilon
{\A} ,\forall i,j \} ,\qquad n=1,2,\cdots .
$$
Thus, an element $\rho \epsilon \Omega^1 ({\A})$ has the form
$$
\rho =\sum_i a^i db^i, \eqno(2.8)
$$
and we impose the condition
$$
\sum_i a^ib^i =1,
$$
since $d1=0 $.
Let $\pi_0 $ denote the representation of ${\A} $ on the space
$h_1\ot h_2$ of square integrable spinors for $SO(1,3)\times SO(10)$,
where $h_2$ is the 32-dimensional Hilbert space on which
${\A}_2$ acts. Let $\overline {\pi_0} $ denote the anti-representation
given by
$$
\overline {\pi_0} (a) =B\overline {\pi_0 (a) }B^{-1} . \eqno(2.9)
$$
We then define $\pi (a) $ by
$$
\pi (a) =\pi_0 (a)+\pi_0 (a)+\overline {\pi_0} (a) \eqno(2.10)
$$
acting on the Hilbert space
$$
\tilde h =h_1 \ot \bigl(h_2^{(1)}\bp h_2^{(2)}\bp h_2^{(3)}
\bigr) ,
$$
where $h_2^{(i)} \cong h_2 ,\qquad i=1,2,3.$
Let $h$ denote the subspace of $\tilde h $ which is the image
of the orthogonal projection onto elements of the form
$$
\pmatrix {P_+\psi \cr P_+\psi \cr P_-\psi^c }
$$
in $\tilde h $. Clearly, $h$ is invariant under $\pi ({\A}) $.
(One can think of $h$ as being a space of sections of a
"vector bundle" over ${\A } $.) On $\tilde h $ we define a self-adjoint
Dirac operator $D$ by setting
$$
D =\pmatrix{\di \ot 1\ot 1 &\g5 \ot M_{12}\ot K_{12} &
\g5 \ot M_{13} \ot K_{13}\cr
\g5 \ot M_{21}\ot K_{21} &\di \ot 1\ot 1&\g5 \ot M_{23}
\cr\g5 \ot M_{31} \ot K_{31}& \g5\ot M_{32}\ot K_{32}  &
\di \ot 1\ot 1 \cr}\eqno(2.11)
$$
where the $K_{mn}$ are $3\times 3$ family-mixing matrices
commuting with the $\pi ({\A})$. We impose the symmetries
$M_{12}=M_{21}={\m}_0 $, $M_{13}=M_{23}={\n}_0$,
$M_{31}=M_{32}={\n}_0^*$, with ${\m}_0={\m}_0^* $. Similar conditions
are imposed  on the matrices $K_{mn}$.
For $D$ to leave the subspace $h$ invariant, ${\m}_0 $ and
${\n}_0 $ must have the form
$$\eqalign{
{\m}_0 &= P_+\bigl( m_0+i m_0^{IJ}\Gamma_{IJ} +m_0^{IJKL}\Gamma_{IJKL}
\bigr)P_+ \cr
{\n}_0 &=P_+\bigl( n_0^I \Gamma_I +n_0^{IJK}\Gamma_{IJK}
+n_0^{IJKLM}\Gamma_{IJKLM}\bigr) P_- \cr}\eqno(2.12)
$$
where
$$ \Gamma_{I_1I_2\cdots I_n}={1\over
n!\ } \Gamma_{[I_1}\Gamma_{I_2}\cdots \Gamma_{I_n]}
$$
are  antisymmetrized products of the gamma
matrices.

Next we define an involutive "representation" $\pi :
\Omega^* ({\A})\leftarrow B(h) $ of $\Omega^* ({\A})$ by
bounded operators on $h$; ($B(h)$ is the algebra of bounded
operators on $h$): We set
$$
\pi_0 (a_0 da_1 da_2 \cdots da_n )=\pi (a_0) [D,\pi (a_1)]
[D,\pi (a_2)]\cdots [D,\pi (a_n)]. \eqno(2.9)
$$
The image of a one-form $\rho $ is
$$
\pi (\rho )=\sum_i a^i [D,b^i], \qquad \sum_i a^ib^i=1 .\eqno(2.14)
$$
{}From now on, we shall write $a^i$ and $b^i$, instead of $\pi
(a^i )$ and $\pi (b^i )$, respectively. Every one-form $\rho $
determines a connection, $\nabla $, on $h$: We set
$$
\nabla =D +\pi (\rho ). \eqno(2.15)
$$
The curvature of $\nabla $ is then given by
$$
\theta =\pi (d\rho )+\pi (\rho^2 ) \eqno(2.16)
$$
where
$$
\pi (d\rho ) =\sum_i [D, \pi (a^i )][D, \pi (b^i )].
$$
It is straightforward to compute $\pi (\rho )$ and one gets [10]
$$
\pi (\rho )=\pmatrix{A &\g5 {\m}  K_{12}&
\g5 {\n} K_{13}\cr \g5 {\m} K_{12} &A &
\g5 {\n} K_{23} \cr\g5 {\n}^*  K_{31}&\g5 {\n}^*  K_{32} &
B \overline A B^{-1}\cr}\eqno(2.17)
$$
where the fields $A$, $\m $ and $\n $  are given in terms
of the $a^i$ and $b^i$ by
$$\eqalign{
A&= P_+(\sum_i a^i \di b^i ) P_+ \cr
{\m }+{\m}_0 &= P_+ (\sum_i a^i {\m}_0 b^i)P_+ \cr
{\n }+{\n}_0 &=P_+(\sum_i a^i {\n}_0
B\overline {b^i}B^{-1})P_- \cr}\eqno(2.18)
$$
We can expand these
fields in terms of the $SO(10)$ Clifford algebra as
follows:
$$\eqalign{
A &=P_+ \bigl( ia +a^{IJ}\Gamma_{IJ} +i a^{IJKL}\Gamma_{IJKL}
\bigr) P_+ \cr
{\m} &= P_+\bigl( m+i m^{IJ}\Gamma_{IJ} +m^{IJKL}\Gamma_{IJKL}
\bigr)P_+ \cr
{\n} &=P_+\bigl( n^I \Gamma_I +n^{IJK}\Gamma_{IJK}
+n^{IJKLM}\Gamma_{IJKLM}\bigr) P_- \cr}
\eqno(2.19)
$$
The self-adjointness condition on
$\pi (\rho ) $ implies, after using the hermiticity of the
$\Gamma_I $ matrices, that all the fields appearing in the expansion
of $A, \m $  are real,because both are self-adjoint,
while those in $\n $ are complex.
 The tracelessness condition on
${\rm tr }(\Gamma_1 \pi (\rho ) )$ where $\Gamma_1 $ is the
grading operator given in the first equation of (2.5). This restricts $a=0$
and then this corresponds to the gauge theory of $SU(16)$.
In this case one must also add mirror fermions to cancel the
anomaly and will not be considered here. We shall require instead
that the gauge fields acting on the first and third copies
have identical components in the Clifford algebra basis. Since
$$
B\overline A B^{-1}=P_-(-ia +a^{IJ}\Gamma_{IJ}-ia^{IJKL}\Gamma_{IJKL})
P_-  \eqno(2.20)
$$
This implies that
$$ \eqalign{
a_{\mu}&=0\cr
a_{\mu}^{IJKL} &=0 \cr}  \eqno(2.21)
$$
The above rquirement can be understoood as the physical condition
that the fermions in the first and third copies will have identical
coupling to the gauge fields.
Then the fermionic action will be given by
$$
(\Psi ,{\cal P} (d+\rho ){\cal P}\Psi )= \int d^4x  {\Psi}^*(x)
{\p}\bigl( D+\pi (\rho ) \bigr){\p} \Psi (x) \eqno(2.22)
$$
where
$$
{\cal P} ={\rm diag}(P_{+},P_{+},P_{-}).
$$
To transform this expression from Euclidean space to Minkowski
space in order to impose the space-time chirality condition,
we have to perform the following substitutions: $\gamma^4
\rightarrow i\gamma^0 $, $\g5 \ra -i\g5 $, $\psi^* \ra
\overline{\psi }$, $\psi^{c*} \ra -\overline {\psi^c}$.
Because of space-time chirality, the  field $\m $
decouples from the fermions. Then this is the field that
must acquire a vacuum expectation value breaking $SO(10)$ at
very large energies. The field $\n $ does couple to fermions
and must acquire expectation values that gives the small fermionic
masses, except for possible large values of the components that
give a mass to the right-handed neutrino.

Now we are ready to write the fermionic action in terms of the
component fields
$$\eqalign{
I_{\rm f}&= \int d^4x \Bigl( 2\ \overline {\psi_+}\Bigr[i(\di
+A )\psi_+ +\g5 ({\n} +{\n}_0 )\psi_+^c K_{13}
 \Bigr] \cr
&\qquad \qquad + \overline{\psi_+^c}\Bigl[ i(\di +A )\psi_+^c +\g5 ({\n}^*
+{\n}_0^* )\psi_+  K_{13}^* \Bigr]\Bigl) \cr}
\eqno(2.23).
$$
where $\psi_+ =P_+\psi $ and by $\s $ chirality is equal to $\psi $.
{}From here on and when convenient we shall denote $\m $ by
$P_+{\m}P_+$ and $\n $ by $P_+{\n}P_-$.
Equation (2.23) can be simplified by using the proporties
of the charge conjugation matrices $B$ and $C$:
$$\eqalign{
B^{-1}\Gamma_I B &=-\Gamma_I^T \cr
C^{-1}\gamma_{\mu} C&=-\gamma_{\mu}^T .\cr}\eqno(2.24)
$$
After rescaling $\psi \rightarrow {1\over \sqrt 3} \psi $
the  action (2.23) simplifies to
$$
I_{\rm f} = \int d^4x \Bigl( \overline { \psi_+ } i (\di + A )\psi_+
  -{1\over \sqrt 3 }\bigl( \psi_+^T B^{-1}C^{-1}( {\n}^*
+{\n}_0^* )\psi_+ K_{13}^* +h.c \bigr) \Bigr) \eqno(2.25)
$$
Thus we have achieved our goal of constructing a Dirac operator
that gives the appropriate interactions of an $\s $ unified
gauge theory.

{\bf\noindent 3. The SO(10) symmetry breaking}
\vskip.2truecm
\noindent
The symmetry breaking pattern that breaks the gauge group $\s $ must be
coded into the Dirac operator $D$. The Higgs fields at
our disposal are $\m $, and $\n $. In terms of $\s $ representations
these are $\u 1,\ \u {45},\ \u {210}$ in $\m $, and complex $
\u {10}, \ \u {120} $ and $\u {126} $ in $\n $.
To be explicit we shall work in a specific $\Gamma
$ matrix representation first introduced by Georgi and
Nanopolous [2].
The $32\times 32 \ \Gamma $ matrices are represented in terms of tensor
products of five sets of Pauli matrices $\sigma_i ,\tau_i ,\eta_i
,\rho_i ,\kappa_i $ where $i=1,2,3$.
To these matrices we assign the following matrices on the tensor
product space:
$$\eqalign {
&\sigma_i \ra 1_2\ot 1_2\ot 1_2\ot 1_2\ot \sigma_i \cr
&\tau_i \ra 1_2\ot 1_2\ot 1_2 \ot \tau_i \ot 1_2 \cr
&\eta_i \ra 1_2\ot 1_2\ot \eta_i \ot 1_2\ot 1_2 \cr
&\rho_i \ra 1_2\ot \rho_i \ot 1_2\ot 1_2 \ot 1_2 \cr
&\kappa_i \ra \kappa_i \ot 1_2 \ot 1_2 \ot 1_2 \ot 1_2 \cr }\eqno(3.1)
$$
The $\Gamma $ matrices are then given by
$$\eqalign{
\Gamma_i &=\kappa_1 \rho_3\eta_i \cr
\Gamma_{i+3}&= \kappa_1 \rho_1 \sigma_i \cr
\Gamma_{i+6}&= \kappa_1 \rho_2\tau_i \cr
\Gamma_0 &=\kappa_2 \cr
\Gamma_{11}&=\kappa_3 \cr}\eqno(3.2)
$$
where $i=1,2,3,$
and when it is obvious we shall
omit the tensor product symbols. In this basis an $\s $ chiral spinor
will take the form
$$
\psi_+ =\pmatrix{\chi_+ \cr 0\cr } \eqno(3.3)
$$
where $\chi $ is a $\u {16}_s $ in the space $V_{\rho} \ot
V_{\eta}
\ot V_{\tau} \ot V_{\sigma}, $ with
$V_{\rho }\equiv \cdots \equiv V_{\sigma}\equiv C^2 $.
The $\s $ conjugation matrix is defined
by $ B\equiv -\Gamma_1\Gamma_3\Gamma_4\Gamma_6\Gamma_8 $ which, in
the basis of equation (3.2), becomes
$$
B=\kappa_1 \rho_2\eta_2\tau_2\sigma_2 \equiv \kappa_1 b \eqno(3.4)
$$
where the matrix $b=\rho_2\eta_2\tau_2\sigma_2 $ is the conjugation
matrix in the space of the
sixteen component spinors. The action
of $B$ on a chiral spinor is then
$$
B\psi_+ =\pmatrix {0 \cr b\chi_+ \cr} \eqno(3.5)
$$
The advantage of this system of matrices is that both
spinors, $\chi_+ $ and
$bC \overline {\chi_+}^T $, have the same form, except for
the first one is left-handed and the second one is right-handed. To correctly
associate the components of $\chi_+ $ with  quarks and leptons,
we consider the action of the charge operator [3] on $\chi_+ $:
$$\eqalign{
Q&={i\over 6} (\Gamma_{45}+\Gamma_{69}+\Gamma_{78} )-{i\over 2}\Gamma_{12}\cr
&=-{1\over 6}(\sigma_3 +\tau_3 +\rho_3\tau_3\sigma_3) +{1\over 2}\eta_3
\cr} \eqno(3.6)
$$
which gives
$$
Q\chi_+ ={\rm diag }(0,{2\over 3},{2\over 3},{2\over
3},-1,-{1\over 3}, -{1\over 3},-{1\over 3}, {1\over 3},{1\over 3},
{1\over 3},1,-{2\over 3},-{2\over 3},-{2\over 3},0)\chi_+ \eqno(3.7)
$$
Thus the components of the left handed spinor $\chi_+ $ are  written as
$$
\chi_+=\pmatrix{n_L \cr u_L^1 \cr u_L^2 \cr
u_L^3 \cr e_L \cr d_L^1 \cr d_L^2 \cr d_L^3 \cr -(d_R^3)^c \cr
(d_R^2)^c \cr (d_R^1)^c \cr -(e_R)^c \cr (u_R^3)^c \cr
-(u_R^2)^c \cr -(u_R^1)^c \cr (n_R)^c \cr }\eqno(3.8)
$$
where the $c$ in this equation stands for the usual charge
conjugation, eg. $d^c =C \overline d^T $. The upper and lower
components in $\chi $ are mirrors, with the signs chosen so
that the spinor $bC\overline {\chi_+}^T $ has exactly the same
form as $\chi_+ $, but with the left-handed and right handed signs,
$L $ and $R$, interchanged.

We now specify the vacuum expectation values (vevs)  ${\m}_0 $ and
${\n}_0 $. The group $\s $ is broken at high energies by
$\m $ which contains the representations $\u {45} $ and
$\u {210} $. By taking the vev of the $\u {210}$ to be
${\m}^{0123}=O(M_G) $, the $\s $ symmetry is broken to $SO(4)\times
SO(6) $ which is isomorphic to $SU(4)_c\times
SU(2)_L\times SU(2)_R $. The $SU(4)_c$ is further broken to
$SU(3)_c\times U(1)_c $ by the vev of the $\u {45}$. Therefore we
write [2-3]
$$\eqalign{
P_+{\m}_0 P_+ &=P_+\Bigl( M_G \Gamma_{0123} -iM_1 (\Gamma_{45}
+\Gamma_{78}+\Gamma_{69} \Bigr)P_+ \cr
&={1\over 2}(1+\kappa_3)\Bigl(- M_G \rho_3 +M_1 (\sigma_3 +\tau_3+
\rho_3\tau_3\sigma_3 ) \Bigr) \cr}\eqno(3.9)
$$
Therefore ${\m}_0 $ breaks $\s $ to $SU(3)_c\times U(1)_c
\times SU(2)_L\times SU(2)_R $ which is also of rank five. The rank
is reduced by giving a vev to the components of
$\u {126} $ that couple to the right-handed neutrino.Therefore
the vev of ${\n}_0$ must contain the term
$$
M_2({1\over 2^5}) (\kappa_1 +i\kappa_2 )(\rho_1 +i\rho_2 )(\eta_1 +i\eta_2 )
(\tau_1 +i\tau_2 )(\sigma_1 +i\sigma_2 ) \eqno(3.10)
$$
In terms of the gamma matrices, equation (3.8) has a rather
complicated form
$$
{1\over 8}\Bigl(\bigl( (\Gamma_{13489}+i (1\ra 2))+i(4\ra 5)\bigr)
-i (8\ra 7)\Bigr). \eqno(3.11)
$$
The vev of ${\n}_0$ break $U(1)_c\times SU(2)_R$ to $U(1)_Y$, and the
surviving group would
be the familiar $SU(3)_c\times SU(2)_L\times U(1)_Y$.
The generators of $SU(2)_L \times SU(2)_R $ are [2]
$$\eqalign{
T_{L,R}^i&=-{i\over 2}({1\over 2}\epsilon^{ijk}\Gamma_{jk}\pm
\Gamma^{i0}) \cr
&={1\over 2}(1\pm \kappa_3\rho_3)\eta^i ,\cr}\eqno(3.12)
$$
while $SU(4)_c$ is generated by
$$\eqalign{
-i\Gamma_{i+3,j+3} &=\epsilon_{ijk}\sigma^k \cr
-i\Gamma_{i+6,j+6} &=\epsilon_{ijk}\tau^k \cr
-i\Gamma_{i+3,j+6} &=\rho_3\tau_j\sigma_i .\cr}\eqno(3.13)
$$
It is  straightforward
to check that the only generators that leave $ {\m}_0 $
and the part of ${\n}_0 $ given by (3.10) invariant
are those of the standard model.
We shall explicitly identify these generators,
in order to proceed to the next stage of breaking
$SU(2)_L \times U(1)_Y $, without any ambiguity. The eight $SU(3)$ generators
are given by $(1-\rho_3\tau_3)\sigma_i $,
$(1-\rho_3\sigma_3)\tau_i $, $\rho_3 (\tau_1\sigma_1 +\tau_2\sigma_2)$
and $\rho_3 (\tau_2\sigma_1 -\tau_1\sigma_2) $. Finally the $U(1)_Y$
generator is
$$
Y=-{1\over 3}(\sigma_3 +\tau_3 +\rho_3\tau_3\sigma_3 )+{1\over 2}
(1-\kappa_3 \rho_3)\eta_3 ,  \eqno(3.14)
$$
and its action on the spinor $\chi_+ $ is given by
$$
Y\chi_+ ={\rm diag }(-1,{1\over 3},{1\over 3},{1\over 3},-1,
{1\over 3},{1\over 3},{1\over 3},{2\over 3},{2\over 3},{2\over 3},
2,-{4\over 3},-{4\over 3},-{4\over 3},0)\chi_+ \eqno(3.15)
$$
This is related to the charge operator $Q$ by
$$
Q={1\over 2} Y+T_L^3  \eqno(3.16)
$$
where the action of the $SU(2)_L$ isospin $T_L^3$ on $\chi_+ $ is given by
$T_L^3={1\over 2}(1+\rho_3)\eta_3 $.

For the last stage of symmetry breaking of $SU(2)_L\times
U(1)_Y$ we can use the field $\n $ which contains the
compex representations $\u {10} $, $\u {120} $ and $\u {126} $. The
most general vev that preserves the group $SU(3)_c \times U(1)_Q
$ is
$$\eqalign{
P_+{\n}_0 P_-&={1\over 2}(1+\kappa_3 )\Bigl( \bigl( is\Gamma_0 +p\Gamma_3
\bigr) \cr
& +\bigl( a'\Gamma_{120}-ia\Gamma_{123} +b'(\Gamma_{453}+\Gamma_{783}
+\Gamma_{693} )-ib(\Gamma_{450}+\Gamma_{690}+\Gamma_{780}) \bigr) \cr
& -\bigl( ie (\Gamma_{01245} +\Gamma_{01269}
+\Gamma_{01278} )+f(\Gamma_{31245}+\Gamma_{31269}+\Gamma_{31278}
)\bigr) \Bigl) \cr
&+{\rm term\ in \ (3.11)}\cr}\eqno(3.17)
$$
Use of the explicit matrix representation for the $\Gamma $ matrices
simplifies eq (3.17) to
$$\eqalign{
P_+{\n}_0 P_-\kappa_1 &={1\over 2}(1+\kappa_3 )\Bigl( s+p\rho_3\eta_3
+a\rho_3 +a'\eta_3 \cr
&\qquad +(b'+b\rho_3\eta_3 +e\eta_3 +f\rho_3 )(\sigma_3 +\tau_3
+\rho_3\tau_3\sigma_3 )\cr
&\qquad +M_2 ({1\over 2^5})(\rho_1 +i\rho_2 )(\eta_1 +i\eta_2 )(\tau_1
+i\tau_2 )(\sigma_1 +i\sigma_2 )
\Bigr), \cr}\eqno(3.18)
$$
where all terms containing $\eta_3 $ break $SU(2)_L \times
U(1)_Y $.
Having specified all the vevs that break $\s $ down to the low-energy
symmetry, it is straightforward, though tedious, to write down
the fermionic masses generated through the symmetry breaking. These
are
$$\eqalign{
I_{{\rm f mass}}&=-{1\over \sqrt 3}\int d^4 x \Bigl(
\bigl[ (s+p +3(e+f))K_{(pq)}
+(a+a'+3(b+b'))K_{[pq]} \bigr] \overline {N_R^p} N_L^q \cr
&\qquad +\bigl[ (s+p-(e+f))K_{(pq)} +(a+a'-(b+b'))K_{[pq]}
\bigr] \overline {u_R^p} u_L^q \cr
&\qquad +\bigl[ (s-p-3(e-f))K_{(pq)} +(a-a'-3(b-b'))K_{[pq]}
\bigr] \overline {e_R^p} e_L^q \cr
&\qquad +\bigl( (s-p +e-f)K_{(pq)} +(a-a'+b-b')K_{[pq]}\bigr)
\overline {d_R^p} d_L^q \cr
&\qquad +\bigl[ M_2 K_{(pq)}(N_R^{pc})^T C^{-1}N_R^{qc}
\bigr] +h.c \Bigr) \cr}\eqno(3.19)
$$
where we have denoted the family mixing matrix $K_{13}$
by $K $. For the neutral fields $N_L$ and $N_R$ we
have a see-saw mechanism giving the right-handed neutrino
a large Majorana mass [11-12],
and the neutrino mass matrix  takes the simple form
(ignoring generation mixing)
$$
\bordermatrix{&N_L&N_R^c \cr N_L&0&m\cr N_R^c&m&M_2\cr}\eqno(3.20)
$$
where $m$ is of order of the weak scale. This matrix
has two eigenstates of masses $M_2$ and ${m^2\over M_2}$.
The free parameters at this stage are $M_G,\ M_1,\ M_2,\
a,\ a',\ b,\ b',\ e,\ f,\ s$ and $p$ and the matrix $K_{pq}$.
However, when we will examine the
scalar potential in the next section, it will become clear that,
in order for the potential, or some terms in it, not to vanish
the above parameters must be related. Also
we note that, since both the symmetric and antisymmetric parts
of $K_{pq}$ enter the fermionic mass matrix, it cannot be
completely removed. By performing a unitary transformation on
$\chi^p_+ \ra U_q^p \chi_+^q $ such that $U^* U=1$ the matrix
$K_{pq}$ is transformed to $(U^TKU)_{pq}$. Since $K$ is an
arbitrary complex matrix, the matrix $U$ can be used to eliminate
nine out of the eighteen real parameters. We shall come back to the
fermionic mass terms after having examined the bosonic sector.

{\bf\noindent 4. The bosonic action  }
\vskip.2truecm
\noindent
In the non-commutative formulation of the Yang-Mills action, an essential
ingredient is the Dirac operator. The curvature of the one-form
$\rho $ is defined by
$$
\theta =d\rho +\rho^2 . \eqno(4.1)
$$
The Yang-Mills action in the non-commutative setting is
given by
$$
I_b ={1\over 4} {\rm Tr}_{\omega} (\theta^2 \vert D \vert^{-4} )\eqno(4.2)
$$
where ${\rm Tr}_{\omega}$ is the Dixmier trace. It was shown in [13]
that one can equivalently use the heat-kernel expression
$$
{\rm lim}_{\epsilon\ra 0}
{{\rm tr}(\theta^2 e^{-\epsilon \vert D
\vert^2 } ) \over
{\rm tr}(e^{-\epsilon \vert D\vert^2 })}\eqno(4.3)
$$
For both definitions, it can be shown that the Yang-Mills action is
equal to [4-5]
$$
I={1\over 4} \int d^4x {\rm Tr}\Bigl( {\rm tr}\bigl( \pi^2 (\theta )
\bigr)\Bigr) \eqno(4.4)
$$
To compute $\pi (\theta  )$, the expression $\pi (d\rho )$ must be
evaluated from the definition of $\rho $:
$$
\pi (d\rho )=\sum_i [D, a^i][D,b^i]  \eqno(4.5)
$$
and this must be expresssed in terms of the fields appearing in
$\pi (\rho )$. Since $\pi (d \rho )$ is not
necessarily zero when $\pi (\rho
)$ is, one must quotient out the space ${\rm Ker}(\pi )+ d{\rm
Ker}(\pi ) $. Since the Yang-Mills action is quadratic in the
curvature $\theta $, the process of working on the quotient space
is equivalent to introducing non-dynamical auxiliary fields
and eliminating them through there equations of motion. The Yang-Mills
action in eq.(4.4) has been derived for an $N$ point space
in [10]. Here we simply quote the result:
$$\eqalignno{
I&=\sum_{m=1}^N Tr\Bigl( {1\over 2}  F_{\mu\nu}^m F^{\mu\nu m}
-\Bigl\vert  \sum_{p\not= m}\vert K_{mp}\vert^2
\vert \phi_{mp}+M_{mp}\vert^2
-(Y_m +X'_{mm})\Bigr\vert^2 \cr
& + \sum_{p\not=m} \vert K_{mp}\vert^2
\Bigl\vert \partial_{\mu}
(\phi_{mp}+M_{mp})+A_{\mu m}(\phi_{mp}+M_{mp})-(\phi_{mp}+M_{mp})
A_{\mu p}\Bigr\vert^2 \cr
& - \sum_{n\not=m}\sum_{p\not=m,n}\Bigl\vert
K_{mp}K_{pn}\bigl(
(\phi_{mp}+M_{mp})(\phi_{pn}+M_{pn})-M_{mp}M_{pn}\bigr) -X_{mn}\Bigr)
\Bigl\vert^2
\Bigr) &(4.6)\cr}
$$
where the $A^m$ are the gauge fields in the $m-m$ entry of $\pi
(\rho )$ and $\phi_{mn} $ are the scalar fields in the $m-n$ entry
of $\pi (\rho )$. The $X_{mn}$, $X'_{mn}$ and $Y_m$ are fields whose
unconstrained elements are auxiliary fields that can be
eliminated from the action. Their expressions in terms of the
$a^i$ and $b^i $ are
$$\eqalignno{
X_{mn}&=\sum_i a_m^i \sum_{p\ne m,n} K_{mp}K_{pn} (M_{mp}M_{pn}
b_n^i -b_m^i M_{mp}M_{pn} ), \qquad m\ne n &(4.7)\cr
X'_{mm}&=\sum_i a_m^i\di^2b_m^i +(\partial^{\mu}
A_{\mu}^m+A^{\mu m}A_{\mu}^m )&(4.8) \cr
Y_m&=\sum_{p\ne m}\sum_i a_m^i\vert K_{mp}\vert^2 \vert M_{mp}
\vert^2 b_m^i &(4.9)\cr}
$$
In the case at hand the discrete space has three points.
Because of the permutation
and complex conjugation symmetry, the $a_m^i$ are related to
each other. This in turn relates some of the auxiliary fields
to one another.
To use eq.(4.5), we must compute the different terms as functionals of
the component fields appearing in $\pi (\rho )$. We first write
$$
A={g\over 4}\gamma^{\mu}A_{\mu}^{IJ}\Gamma_{IJ}
$$
where $g$ is the $\s $ gauge coupling constant. Then the kinetic term
for the gauge field $A_{\mu}^{IJ}$ as given by the first term in
eq (4.6), after computing the sum and the trace over  $\cl $,
is equal to
$$
-4g^2 F_{\mu\nu}^{IJ}F^{\mu\nu IJ} \eqno(4.10)
$$
where the field strength is
$$
F_{\mu\nu}^{IJ} =\partial_{\mu}A_{\nu}^{IJ}-\partial_{\nu}A_{\mu}^{IJ}
+g(A_{\mu}^{IK}A_{\nu}^{KJ}-A_{\nu}^{IK}A_{\mu}^{KJ}) \eqno(4.11)
$$
The Higgs kinetic terms have two parts, corresponding to $\m $
and $\n $. Using the decompositon of $\m $ and $\n $ in the
$\cl $-basis one gets the result
$$\eqalign{
&64{\rm Tr}\vert K_{12}\vert^2 \Bigl( \bigl( \partial_{\mu}m)^2\bigr)
+2 \bigl( D_{\mu}(m+m_0)_{IJ}\bigr)^2 +4\bigl( D_{\mu}(m+m_0)_{IJKL}\bigr)^2
\Bigr) \cr
&+64\vert K_{13}\vert^2 \Bigl( \vert D_{\mu}(n+n_0)_I\vert^2
+3\vert D_{\mu}(n+n_0)_{IJK}\vert^2 +5\vert D_{\mu}(n+n_0)_{IJKLM}
\vert^2 \Bigr) \cr}\eqno(4.12)
$$
where the $D$ appearing in this equation is the covariant derivative
with respect to the $\s $ gauge group, and the $m $'s and $n$'s
are defined in eq. (2.19). For example
$D_{\mu} n_I =\partial_{\mu}n_I +g A_{\mu}^{IJ}n_J $. The masses
of the components of the gauge fields $A_{\mu}^{IJ} $ corresponding
to the broken generators of $\s $ are provided by the vevs ${\m}_0
$, and ${\n}_0$ . The most complicated part is the Higgs potential,
since this involves new fields  some of which are related, and
the non-dynamical ones must be eliminated through their equations of motion.
It is given by
$$\eqalign{
V({\m },{\n })&=2\Bigl\vert \vert K_{12}\vert^2 \vert {\m }+{\m }_0\vert^2
+\vert K_{13}\vert^2 \vert {\n}+{\n }_0 \vert^2
-(Y_1+X'_{11})\Bigr\vert^2\cr
& +\Bigl\vert \vert K_{31}\vert^2\vert {\n}+{\n}_0\vert^2 +\vert K_{12}
\vert^2 \vert {\m}+{\m}_0 \vert^2
 -(Y_3+X'_{33})\Bigr\vert^2 \cr
& +2\Bigl\vert \vert K_{13}\vert^2 \vert ({\n}+{\n}_0 \vert^2
-\vert {\n}_0 \vert^2 ) \vert^2
-X_{12}\Bigr\vert^2 \cr
& +2\Bigl\vert K_{12}K_{23}\bigl( ({\m} +{\m}_0 )({\n} +{\n}_0)-{\m}_0
{\n}_0 \bigr)  -X_{13}\Bigr\vert^2 ,\cr}\eqno(4.13)
$$
where we have used the symmetry that equates some of the $K's$
and of the $X's$. We now write the explicit expressions for the $X$
and $Y$ fields. First, we have :
$$\eqalign{
X'_{11}&=\sum_i a^i \partial^2 b^i +(\partial^{\mu} A_{\mu} +A^{\mu}
A_{\mu})\cr
X'_{33}&= B\overline {X'_{11}}B^{-1} \cr}\eqno(4.14)
$$
Next, we have for the $Y$'s
$$\eqalign{
Y_1&=\sum_i a^i \vert K_{12}\vert^2 \vert {\m}_0\vert^2 b^i
+2\sum_i a^i \vert K_{13}\vert^2 \vert {\n}_0\vert^2 b^i\cr
Y_3&=B\overline{ Y_1}B^{-1} \cr}\eqno(4.15)
$$
Finally we have for the $X_{mn}, m\ne n$, the expressions
$$\eqalign{
X_{12}&= \vert K_{13} \vert^2 \Bigl( \sum_i a^i \vert {\n}_0 \vert^2 B
\overline {b^i} B^{-1} -\vert {\n}_0\vert^2 \Bigr) \cr
X_{13}&=K_{12}K_{23}\Bigl( \sum_i a^i {\m}_0{\n}_0B\overline {b^i}B^{-1}
-{\m}_0{\n}_0 \Bigr),  \cr}\eqno(4.16)
$$
and the other $X$'s are related to the above ones by permutation symmetry.
It is easy to notice that $X'_{11}$ and $X'_{33}$ are auxiliary
fields that do not depend on the $K$ matrices. Therefore,
eliminating these fields would result in expressions orthogonal
to the corresponding $K$ space. Eliminating the remaining
auxiliary fields $Y_1,\ Y_3,\  X_{12}$, and $ X_{13}$ is much more
complicated. If all of these were independent the potential
would vanish, after eliminating them. However, if the vevs
${\m}_0 $ and ${\n}_0 $ are chosen in a special way then it
is possible for the potential to survive.
One must arrange for a relation  between
the auxiliary fields, so that, after eliminating the independent
combinations, the potential that corresponds to the given vacuum
will result.  A close look at the potential in eq. (4.13) shows that
if all of the $X$ and $Y$ fields are independent, the  potential
disappears after eliminating them. By comparing $X_{12}$ and
$Y_1$ one sees that they can be related only if $\sum a^i \vert
{\m}_0\vert^2 b^i$ is not an independent field. This can happen
if
$$
M_G=M_1 \eqno(4.17)
$$
so that $\vert {\m}_0 \vert^2 =4M_1^2 $, and we get the relation
$$
Y_1=\vert K_{12}\vert^2 \vert {\m}_0 \vert^2 +\vert K_{13}\vert^2
\vert {\n}_0 \vert^2 +X_{12}\eqno(4.18)
$$
Next, for the term in the potential depending on $X_{13}$ not to
vanish, $X_{13}$ must not be an independent field and must be
a function of ${\n}$. This is possible if ${\m}_0{\n}_0$ is proportional
to ${\n}_0$. This condition is extremely restrictive, but fortunately
has one solution given by
$$
\eqalign{
{\m}_0{\n}_0&=2M_1{\n}_0 \cr
a'&=b'=0\cr
f&=-s={a\over 2} \cr
p&=3e={3\over 2}b \cr}\eqno(4.19)
$$
and the free parameters in the theory are $M_1,\ M_2,\ a,\ b$ and
the matrices $K_{12},\ K_{13}$. The equation for $X_{13}$ simplifies
to
$$
X_{13}=K_{13}(2M_1 {\n})\eqno(4.20)
$$
Then the only independent fields to be eliminated are $X_{12}$ and
$X_{11}^{'}$. The resulting potential is
$$\eqalign{
V({\m},{\n})&=\bigl( {\rm Tr}\vert K_{12}\vert^4 -({\rm Tr}\vert
K_{12}\vert^2)^2\bigr) \Bigl\vert \vert {\m}+{\m}_0 \vert^2
-4M_1^2\Bigr\vert^2 \cr
&+2{\rm Tr}\vert K_{12}K_{13}\vert^2 \Bigl\vert ({\m}+{\m}_0-2M_1)
({\n}+{\n}_0)\Bigr\vert^2\cr}\eqno(4.21)
$$
The total bosonic action is the sum of the terms (4.10), (4.12)
and (4.21), multiplied by an overall constant. We choose this constant
to be ${1\over 16g^2}$ to get the canonical kinetic energy for the
gauge fields. The kinetic energy for the scalar fields, ${\m}$ and
${\n}$, is normalized canonically after rescaling
$$\eqalign{
{\m}&\ra {g\over 2\sqrt {2{\rm Tr}\vert K'\vert^2}}{\m} \cr
{\n}&\ra {g\over 2\sqrt {{\rm Tr}\vert K \vert^2 }}{\n},\cr}\eqno(4.22)
$$
where we have denoted $K_{13}$ by $K$ and $K_{12}$ by $K'$.
After rescaling, the bosonic action becomes
$$\eqalign{
I_{\rm bosonic}&=\int d^4x\Bigl(-{1\over 4}F_{\mu\nu}^{IJ}F^{\mu\nu IJ} \cr
&+{1\over 32}{\rm Tr}\Bigl[ {1\over 2}\bigl( D_{\mu} ({\m}+{\m}_0)\bigr)^2
+\bigl\vert D_{\mu}({\n}+{\n}_0)\bigr\vert^2\Bigr]\cr
&+{g^2 \over 2^5.32}\Bigl( {{\rm Tr}\vert K'\vert^4\over ({\rm Tr}\vert
K'\vert^2 )^2}-1\Bigr) {\rm Tr}\Bigl\vert \vert {\m}+{\m}_0\vert^2
-4M_1^2\Bigr\vert^2\cr
&+{g^2\over 2^3.32}\Bigl\vert ({\m}+{\m}_0-2M_1)({\n}+{\n}_0)\Bigr\vert^2
\Bigr) \cr}\eqno(4.23)
$$
Finally, the fermionic action becomes
$$\eqalign{
I_{\rm f}&=-{g\over \sqrt {3{\rm Tr}\vert K\vert^2}}
\int d^4x \Bigl( K_{pq}\bigl(
(a+3b)\overline {N_R^p}N_L^q +(a-3b)\overline {e_R^p}e_L^q\bigr)
\cr &\qquad +K_{qp}\bigl( (-a+b)\overline {u_R^p}u_L^q -(a+b)
\overline {d_R^p}d_L^q\bigr) +M_2 K_{(pq)}(N_R^p)^TC^{-1}N_R^q
+h.c \Bigr)\cr }\eqno(4.24)
$$
By examining the gauge kinetic term one finds the usual $ \s $ relations
among the gauge coupling constants
$$
g_2=g_3=g=\sqrt {{5\over 3}}g_1 \eqno(4.25)
$$
implying that $\sin^2 \theta_{W} $, at the unification scale
$M_1$ is ${3\over 8 }$. From the ${\n} $-kinetic term one sees that
the mass of the W gauge boson is
$$
m^2_W ={g^2\over 4}(a^2+3b^2) \eqno(4.26)
$$
{}From the fermionic mass terms, one deduces,
using the fact that the top
quark mass is much heavier than the other fermionic masses,
that
$$
m_t=g\vert b-a \vert \eqno(4.27)
$$
Comparing with $m_W $ we get the relation
$$
m_t =2m_W{\vert 1-{b\over a}\vert\over \sqrt {1+{3b^2\over a^2}}}\eqno(4.28)
$$
and this gives  upper and lower bounds on the top quark mass
$$
{2\over \sqrt 3}m_w\le m_t\le {4\over \sqrt 3}m_w=186.13 {\rm Gev} \eqno(4.29)
$$
which agrees with  present experimental limits.
Unfortunately, the same matrix $K_{qp}$ appears for the $u^p$
and $d^p$ quarks, implying that the same transformation can be
used for $u^p$ and $d^p$ to diagonalize $K_{qp}$. This in turn
implies that this model does not allow for a Cabibbo angle, and
this phenomenologically rules out this model.  This forces us
to look for modifications in this model so that it becomes
acceptable.  This result shows that model building in  non-commutative
geometry is so constrained that the models could be ruled out
on phenomenological grounds.

{\bf \noindent 5. A realistic SO(10) model }
\vskip.2truecm
\noindent
The model presented in the previous sections is minimal in the sense
that the number of  points in the internal geometry and the Higgs
fields cannot be reduced. If one insists on a $Z_2$ symmetry between the
different copies, then the number of points would have to be even, and we
have to take two copies where the conjugate spinors are placed, instead
of the one copy considered before. It will be seen that this extension cannot
have a potential after eliminating the auxiliary fields.
Therefore, this model has
to be further extended by one or two points to get the
 $\underline {16}_{\rm s}$  Higgs field, and this will ensure
that the potential can be arranged to survive.
The fermionic space is extended with a singlet
spinor. Two of the neutral fermions will become superheavy,
while the third one would  remain massless.
The triple $({\A},h,D)$ is defined in the same way as in
section 2, with the algebra  ${\A}_2 $ given by
$$
{\A}_2 \equiv P_+ {\cl}P_+ \bp R , \eqno(5.1)
$$

The involutive map $\pi $ is now taken
to be:
$$
\pi (a) =\pi_0 (a) \bp \pi_0 (a) \bp \overline {\pi}_0 (a)
\bp \overline {\pi}_0 (a) \bp \pi_1 (a)\bp \pi_1 (a)  \eqno(5.2)
$$
acting on the Hilbert space
$$
\tilde h =h_1\ot (h_2^{(1)}\bp \cdots \bp h_2^{(6)}) \eqno(5.3)
$$
where $h_2^{(i)}\cong h_2, \qquad i=1 \cdots 4,$ and $h_2^{(i)}\cong C
\qquad i=5,6$. Let $h$ denote the subspace of $\tilde h$ which
is the image of the orthogonal projection into elements of the form
$$
\Psi \equiv \pmatrix {P_+\psi \cr P_+\psi \cr P_-\psi^c \cr
P_-\psi^c
 \cr \lambda \cr \lambda^c \cr}, \eqno(5.4)
$$
On $\tilde h$ the self-adjoint Dirac operator $D$ becomes
$$
D =\pmatrix{\di \ot 1\ot 1 &\g5 \ot M_{12}\ot K_{12} &
\ldots &\g5 \ot M_{16} \ot K_{16}\cr
\g5 \ot M_{21}\ot K_{21} &\di \ot 1\ot 1&\ldots &\g5 \ot M_{26}
\cr \vdots &\vdots &\ddots &\vdots \cr
\g5 \ot M_{61} \ot K_{61}& \g5\ot M_{62}\ot K_{62} &\ldots &
\di \ot 1 \cr}\eqno(5.5)
$$
where the $K_{mn}$ are $3\times 3$ matrices commuting with the $a_i$
and $b_i$.
Therefore we shall take
$$\eqalign{
M_{12}&=M_{21}={\m}_0\cr
M_{34}&=M_{43}=BM_{12}B^{-1}\cr
M_{13}&=M_{23}=M_{14}=M_{24}={\n}_0\cr
M_{15}&=M_{16}=M_{25}=M_{26}=H_0 \cr
M_{35}&=M_{45}=M_{36}=M_{46}=BM_{15} \cr
M_{56}&=0 \cr }\eqno(5.6)
$$
where ${\m}_0$ and ${\n}_0$ are given by eq. (2.12).
Similar symmetry conditions are imposed on  $K_{mn}$.
For $\pi (\rho )$  one  then gets
$$
\pi (\rho )=\pmatrix{A &\g5 {\m}  K_{12}&
\g5 {\n} K_{13}&\g5 {\n} K_{14}&\g5 H K_{15}&
\g5 H K_{16}\cr \g5 {\m} K_{12} &A &
\g5 {\n} K_{23} &\g5 {\n} K_{24}&\g5  H K_{25}
&\g5  H  K_{26}\cr
\g5 {\n}^*  K_{31}&\g5 {\n}^*  K_{32} &B \overline A B^{-1}
 &\g5 {\m }{'} K_{34} &
\g5 H' K_{35}&\g5 H'  K_{36}\cr
\g5 {\n}^*  K_{41}&\g5 {\n}^* K_{42}&\g5
{\m }^{'} K_{43}&B \overline A B^{-1}  &\g5 H'  K_{45}&\g5  H' K_{46}\cr
\g5 H^* K_{51}&\g5 H^* K_{52}&\g5 H^{'*} K_{53}
&\g5 H^{'*} K_{54}&0&0\cr
\g5 H^*  K_{61}&\g5 H^* K_{62}&\g5 H^{'*} K_{63}&
\g5 H^{'*} K_{64}&0& 0\cr}\eqno(5.7)
$$
where the new functions $A$, $\m $, $\n $ and $H$ are given in terms
of the $a^i$ and $b^i$ by
$$\eqalign{
A&=P_+(\sum_i a^i \di b^i )P_+ \cr
{\m}+{\m}_0 &=P_+(\sum_i a^i {\m}_0 b^i )P_+ \cr
{\n}+{\n}_0 &=P_+(\sum_i a^i {\n}_0 B\overline {b^i}B^{-1})P_- \cr
H+H_0 &=P_+(\sum_i a^i H_0 b^{'i})\cr}\eqno(5.8)
$$
and
$$
{\m}^{'}=B\overline {\m} B^{-1}, \eqno(5.9)
$$
$$H'=B\overline H.  \eqno(5.10)
$$
We shall make the same physical requirement as in eqs. (2.20)
and (2.21) that reduces the
gauge group from $U(16)$ to $SO(10)$.
The fermionic action, in terms of the
component fields, is given by
$$\eqalign{
I_f&= \int d^4x \Bigl( 2\ \overline {\psi_+}\Bigr[i(\di
+A )\psi_+ +2\g5 ({\n} +{\n}_0 )\psi_+^c K_{13}+\g5 (
H+H_0) \lambda^c K_{15} \Bigr] \cr
&\qquad -2\ \overline{\psi_+^c}\Bigl[ i(\di +A )\psi_+^c +2\g5 ({\n}^*
+{\n}_0^* )\psi_+  K_{13}^*\g5 B(\overline H+\overline {H_0})\lambda
K_{15}^* \Bigr] \cr
&\qquad +\overline{\lambda }\Bigl[ i\di \lambda +2\g5 (H+H_0)^T B^{-1}
\psi_+^c K_{15} + \Bigr]  -\overline{\lambda ^c} \Bigl[ i\di \lambda^c +2\g5
(H^* +H_0^*)\psi_+ K_{15}^*  \Bigr]\Bigl)}
\eqno(5.11).
$$
where $\psi_+ =P_+\psi $ and by $\s $  chirality is equal to $\psi $.
This expression can be simplified by using the proporties
of the charge conjugation matrices $B$ and $C$ and,
after rescaling $\psi \rightarrow {1\over 2}\psi $ and $\lambda \rightarrow
{1\over \sqrt 2}\lambda $, the fermionic action (5.11) simplifies
to
$$\eqalign{
I_f&= \int d^4x \Bigl(\overline{ \psi_+}i(\di +A)\psi_+
+\overline {\lambda} i\di \lambda \cr
&\qquad \quad -\Bigl[ \psi_+^T B^{-1}C^{-1}({\n}^* +{\n}_0^*)
\psi_+K_{13}^* \cr
&\qquad \quad +{1\over \sqrt 2} \lambda^T C^{-1} (H^*+H_0^*)\psi_+ K_{15}^*
+{1\over \sqrt 2} \psi_+^T C^{-1}(\overline H +\overline {H_0})
\lambda K_{35} +h.c \Bigr] \Bigl) \cr}\eqno(5.12)
$$
The only change in the breaking mechanism is that  $U(1)_c
\times SU(2)_R $ is broken also by the $H_0$ whose
vev  is given by
$$
H_0 =M_3\pmatrix{0\cr\vdots \cr 0\cr 1\cr}\eqno(5.13)
$$
The fermionic action is modified slightly from eq (3.19) to become
$$\eqalign{
I_{{\rm f-mass}}&=-\int d^4 x \Bigl( \bigl( (s+p +3(e+f))K_{(pq)}
+(a+a'+3(b+b'))K_{[pq]} \bigr) \overline {N_R^p} N_L^q \cr
&\qquad +\bigl( (s+p-(e+f))K_{(pq)} +(a+a'-(b+b'))K_{[pq]}
\bigr) \overline {u_R^p} u_L^q \cr
&\qquad +\bigl( (s-p-3(e-f))K_{(pq)} +(a-a'-3(b-b'))K_{[pq]}
\bigr) \overline {e_R^p} e_L^q \cr
&\qquad +\bigl( (s-p +e-f)K_{(pq)} +(a-a'+b-b')K_{[pq]}\bigr)
\overline {d_R^p} d_L^q \cr
&\qquad +\bigl( \sqrt 2 M_3 K_{pq}^{'}\overline {N_R^p}\lambda_L^q
+ M_2 K_{(pq)}(N_R^{pc})^TC^{-1}N_R^{qc}\bigr)
+h.c \Bigr) \cr}\eqno(5.14)
$$
where we have denoted the family mixing matrices $K_{13}$, $K_{15}$
and $K_{56}$  by $K, K', K^{''} $, respectively.
Since we have three
neutral fields, $N_L$, $N_R^c $ and $\lambda_L$, and their mass eigenstates
are mixed, the mass matrix must be diagonalised. Ignoring the
mixing due to the generation matrices, the mass matix is of the form
$$
\bordermatrix{&N_L&N_R^c&\lambda_L \cr
N_L&0&m&0\cr N_R^c &m&M_2&M_3 \cr \lambda_L &0&M_3&0\cr}\eqno(5.15)
$$
and we shall assume a mass hierarchy $m\ll M_2,M_3$, and
$M_2\sim M_3$.
Diagonalisation of the matrix (5.13) produces two massive fields
whose masses are of order $M_2$, and the third will be a massless
left-handed neutrino.
The kinetic term for the gauge field $A_{\mu}^{IJ}$
is equal to
$$
-4g^2 F_{\mu\nu}^{IJ}F^{\mu\nu IJ} \eqno(5.16)
$$
and the Higgs kinetic terms have three parts corresponding to $\m $,
$\n $ and $H$. They are given by
$$\eqalign{
&2\vert K_{12}\vert^2 {\rm Tr}\Bigl(
\bigl( D_{\mu}({\m}+{\m}_0)\bigr)^2 \Bigr)
 +8\vert K_{13}\vert^2 {\rm Tr}\Bigl( \vert D_{\mu}({\n}+{\n}_0)
\vert^2 \Bigr) \cr
& +12\vert K_{15}\vert^2 \Bigl\vert D_{\mu}(H+H_0) \Bigr\vert^2 \cr}\eqno(5.17)
$$
where the $D$ appearing in this equation is the covariant derivative
with respect to the $\s $ gauge group. The mass terms of the
gauge fields corresponding
to the broken generators of $\s $ are provided by the vevs ${\m}_0
$, ${\n}_0$ and $H_0$. The Higgs potential is very complicated in this
case. It is given by
$$\eqalign{
& 2\Bigl\vert \vert K_{12}\vert^2 \vert {\m }+{\m }_0\vert^2
+2\vert K_{13}\vert^2 \vert {\n}+{\n }_0 \vert^2
+2\vert K_{15}\vert^2 \vert H+H_0 \vert^2 -(Y_1+X'_{11})\Bigr\vert^2\cr
& +2\Bigl\vert 2\vert K_{31}\vert^2\vert {\n}+{\n}_0\vert^2 +\vert K_{12}
\vert^2 \vert {\m}+{\m}_0 \vert^2 +2\vert K_{13}\vert^2 \vert H+H_0
\vert^2 -(Y_3+X'_{33})\Bigr\vert^2 \cr
& +2\Bigl\vert 4\vert K_{51}\vert^2 H+H_0 \vert^2 -(Y_5 +X'_{55}) \Bigr\vert^2
\cr & +2\Bigl\vert 2\vert K_{13}\vert^2 \vert ({\n}+{\n}_0 \vert^2
-\vert {\n}_0 \vert^2 ) +2 \vert K_{15}\vert^2 (\vert H+H_0\vert^2
-\vert H_0\vert^2 )-X_{12}\Bigr\vert^2 \cr
& +8\Bigl\vert K_{12}K_{23}\bigl( ({\m} +{\m}_0 )({\n} +{\n}_0)-{\m}_0
{\n}_0 \bigr) \cr
& \qquad +K_{14}K_{43}\bigl( ({\n}+{\n}_0 )(\overline {{\m}}
+\overline {{\m}_0})-{\n}_0 \overline {{\m}_0} \bigr)
\cr &\qquad +2K_{15}K_{53}\bigl( (H+H_0)B(\overline H +\overline
{H_0} -H_0B\overline {H_0}\bigr) -X_{13}\Bigr\vert^2 \cr
&+8\Bigl\vert K_{12}K_{25}({\m} +{\m}_0)(H+H_0)+2K_{13}K_{35}({\n}+{\n}_0)
B(\overline H+\overline {H_0})  -X_{15}\Bigr\vert^2 \cr
&+2\Bigl\vert 2K_{31}K_{14}\bigl( \vert {\n}^* +{\n}_0^*\vert^2 -\vert
{\n}_0^*\vert^2 \bigr) +2\vert K_{35}\vert^2 \bigl( \vert
B(\overline H +\overline {H_0} )\vert^2 -\vert B\overline {H_0}\vert^2
\bigr) -X_{34}\Bigr\vert^2 \cr
&+8\Bigl\vert 2K_{31}K_{15}\bigl(
({\n}^*+{\n}_0^*)(H+H_0)-{\n}_0^*H_0^* \bigr) \cr
&\qquad +2K_{34}K_{45}
\bigl( \vert B(\overline H+\overline {H_0})\vert^2 -\vert B\overline
{H_0}\vert^2 \bigr) -X_{35}\Bigr\vert^2 \cr
& +2\Bigl\vert 4\vert K_{51}\vert^2 \bigl( \vert H^*+H_0^*\vert^2 -\vert
H_0^*\vert^2 \bigr) -X_{56}\Bigr\vert^2 \cr}\eqno(5.18)
$$-
where we have used the symmetry that equates some of the $K's$
and the $X's$. The explicit expressions for the $X$
and $Y$ fields are:
$$\eqalign{
X'_{11}&=\sum_i a^i \partial^2 b^i +(\partial^{\mu} A_{\mu} +A^{\mu}
A_{\mu})\cr
X'_{33}&= B\overline {X'_{11}}B^{-1} \cr
X'_{55}&=\sum_i a^{'i}\partial^2 b^{'i} \cr}\eqno(5.19)
$$
Next, we have
$$\eqalign{
Y_1&=\sum_i a^i \vert K_{12}\vert^2 \vert {\m}_0\vert^2 b^i
+2\sum_i a^i \vert K_{13}\vert^2 \vert {\n}_0\vert^2 b^i
 +\sum_i a^i \vert K_{15}\vert^2\vert H_0\vert^2 b^i \cr
Y_3&=B\overline{ Y_1}B^{-1} \cr
Y_5&=2M_2^2 (\vert K_{51}\vert^2 +\vert K_{53}\vert^2 )\cr}\eqno(5.20)
$$
The expressions for $X_{mn}, m\ne n$ are now given by
$$\eqalign{
X_{12}&= 2\vert K_{13} \vert^2 \bigl( \sum_i a^i \vert {\n}_0 \vert^2 B
\overline {b^i} B^{-1} -\vert {\n}_0\vert^2 \bigr) +2\vert K_{15}
\vert^2 \bigl( \sum_i a^i \vert H_0 \vert^2 b^i -\vert
H_0\vert^2 \bigr) \cr
X_{13}&=K_{12}K_{23}\bigl(\sum_i a^i {\m}_0{\n}_0B\overline {b^i}B^{-1}
-{\m}_0{\n}_0 \bigr) \cr
&\qquad +K_{14}K_{43}\sum_i a^i {\n}_0B\overline {{\m}_0}B^{-1} -{\n}_0
B\overline{{\m}_0}B^{-1} \bigr)\cr
&\qquad +2\vert K_{15}\vert^2\bigl( \sum_i a^i
H_0\overline{H_0}\overline{b^i}B^
{-1}
-\vert H_0\vert^2 B^{-1} \bigr) \cr
X_{15}&=\vert K_{12}K_{25}\vert \bigl( \sum_i a^i {\m}_0H_0
b^{'i}-{\m}_0H_0 \bigr) +2K_{13}K_{35}\bigl( \sum_i a^i{\n}_0B\overline
{H_0}b^{'i}-{\n}_0B\overline{H_0}\bigr) \cr
X_{34}&=B\overline {X_{12}}B^{-1} \cr
X_{35}&=B\overline{X_{15}}\cr
X_{56}&=0 \cr}\eqno(5.21)
$$
and  the other $X$'s are related to the ones above by permutation
symmetry $X_{12}=X_{21}$, $X_{34}=X_{43}$, $X_{13}=X_{14}=X_{23}
=X_{24}$, $X_{16}=X_{26}$ and $X_{36}=X_{46}$. We also have
similar identities for the $K$'s, and, in addition, we have assumed
the relations
$K_{12}=\overline {K_{34}}$ and $K_{15}=\overline {K_{35}}$.
In analogy with the previous model, we must impose the relation
$$
M_G=M_1,  \eqno(5.22)
$$
in order to get a relation between $X_{12}$ and $Y_1$:
$$
Y_1=\vert K_{12}\vert^2
\vert {\m}_0 \vert^2 +2\vert K_{13}\vert^2 \vert {\n}_0\vert^2
+X_{12} \eqno(5.23)
$$
For the term in the potential involving $X_{13}$ to survive,
we should be able to express this field in terms of the other
scalar fields. By examining the expression for $X_{13}$ we notice
that a simplification occurs if we require that
$$
K_{12}=\overline {K_{12}} \eqno(5.24)
$$
because the terms involving ${\m}_0{\n}_0^{(1)} $ drop out, where
${\n}_0^{(1)} $ is the part of ${\n}_0$ independent of $M_2$. In
this case $X_{13}$ can be made to be zero, provided that we take
$$
M_1M_2=-{K_{15}\overline {K_{15}}\over 2K_{12}K_{13}} M_3^2 \eqno(5.25)
$$
where we have used the relation
$$
H_0H_0^* ={M_2\over M_3^2}{\n}_0^{(2)} \eqno(5.26)
$$
where ${\n}_0^{(2)}$ is the part in ${\n}_0$ dependent on $M_2$.
If no relation is taken between $K_{12}$ and $\overline {K_{12}}$
then the only way for the potential to survive is to impose a
relation on ${\m}_0{\n}_0$ identical to the one for the simpler
model as well as a relation between $M_2$ and $M_3$. This case will
not be interesting for us since the fermionic mass matrices, apart
from the neutral fields sector, are identical to those in the previous
model and thus would suffer the same problem of the absence of the
Cabibbo angle. The auxiliary field $X_{15}$ is not independent and is
equal to
$$
X_{15}=uH,  \eqno(5.27)
$$
where
$$\eqalign{
u&= 2K_{13}\overline {K_{15}}\Bigl(s+p-3(b+b')+2(a+a')+M_2\Bigr)
 -2K_{12}K_{25}M_1 \cr}\eqno(5.28)
$$
After eliminating the auxiliary fields $Y_1$ and $X_{33}^{'}$
the potential becomes
$$\eqalign{
V({\m},{\n},H)&= \bigl( {\rm Tr}\vert K_{12}
\vert^4-({\rm Tr}\vert K_{12}\vert^2)^2
\bigr) {\rm Tr}\Bigl( \vert {\m}+{\m}_0\vert^2 -\vert {\m}_0\vert^2
\Bigr)^2 \cr
&+4\Bigl\vert K_{13}K_{12}\bigl( ({\m}+{\m}_0)({\n}+{\n}_0)
+({\n}+{\n}_0)B(\overline {{\m}}+\overline {{\m}_0})B^{-1}
\bigr) \cr
&\qquad +2K_{15}\overline {K_{15}}\bigl( (H+H_0)B(\overline
H+\overline {H_0})\bigr) \Bigr\vert^2 \cr
& +8\Bigl\vert K_{12}K_{15} ({\m}+{\m}_0)(H+H_0)
+2K_{13}\overline {K_{15}}({\n}+{\n}_0)B(\overline H +\overline {H_0}
)\cr
&\qquad  -u(H+H_0)\Bigr\vert^2\cr
&+16\bigl( {\rm Tr}\vert K_{15}\vert^4 -({\rm Tr}\vert K_{15}\vert^2
)^2\Bigl\vert \vert H^* +H_0^*\vert^2 -M_3^2\Bigr\vert^2\cr
&+16{\rm Tr}\vert K_{15}\vert^4 \Bigl\vert \vert H^* +H_0^*\vert^2
-M_3^2\Bigr\vert^2 \cr}\eqno(5.29)
$$
Therefore the fermionic mass terms are still given by
eq.(5.14) and do not suffer the problem encountered before
This completes our study of the model and shows that it is possible
to obtain a nice $\s $ model. A complete phenomenological analysis
will be left for the future.
\vskip.2truecm
{\bf \noindent 6.Summary and conclusion }
\vskip.2truecm
\noindent
We have seen that a realistic $\s $ model can be constructed using
the non-commutative geometry setting of Connes. The attractiveness
of this model stems from the fact that all the fermions fit into
one representation making the spinor space particularly simple.
Depending on the number of discrete points extending the continuous
geometry the Higgs structure is predicted uniquely. We found two
models: The first one is quite simple and has a very
restrictive form for the fermion masses, but turns out to
be unrealistic. The second example is more
complicated, but the Higgs structure is essentially the same as that
of the first model, with the difference of an additional
$\underline {16_s} $ Higgs field. The fermionic masses  are not
as restricted as those in the first model. We hope to study the spectrum
in more detail, in the future.
A study of the
quantum system is not meaningful before having determined those
symmetries of the system that are characteristic of the
non-commutative geometry setting.

\vskip1truecm
{\bf\noindent Acknowledgments}\hfill\break
We would like to thank D. Wyler for very useful discussions.
\vfill
\eject

{\bf \noindent References}
\vskip.2truecm

\item{[1]} H. Georgi and S. Glashow, {\sl Phys. Rev. Lett.} {\bf 32}
(1974) 438.

\item{[2]} H. Georgi in {\sl Particles and Fields} {\bf 1974} p.575,
editor C.E. Carlson (AIP, New York, 1975);\br
H. Fritzch and P. Minkowski, {\sl Ann. Phys.} (N.Y) {\bf
93}  (1977) 193;\br
M.S. Chanowitz, J. Ellis and M. Gaillard, {\sl Nucl. Phys.} {\bf
B129} (1977) 506.

\item{[3]} S. Rajpoot, Phys. Rev. {\bf D22} (1980) 2244;\br
F. Del Aguila and L. Ibanez, {\sl Nucl. Phys.} {\bf B177} 60 (1981);\br
R. N. Mohapatra and G. Senjanovic, {\sl Phys. Rev.} {\bf D27} (1983)
1601.

\item{[4]} A. Connes, {\sl Publ. Math. IHES.} {\bf 62} (1983) 44.

\item{[5]} A. Connes, in {\sl the interface of mathematics
and particle physics }, Clarendon press, Oxford 1990, Eds
D. Quillen, G. Segal and  S. Tsou

\item{[6]} A. Connes and J. Lott,{\sl Nucl. Phys. Proc. Supp.}
{\bf 18B} (1990) 29, North-Holland, Amsterdam.

\item{[7]} A. Connes and J. Lott, {\sl  Proceedings of
1991  Cargese summer conference}. See also A. Connes
{\sl Non-Commutative Geometry} expanded English translation (to appear)
of {\sl Geometrie Non-Commutative}, Paris, Interedition.

\item{[8]} D. Kastler , Marseille preprints CPT-91/P.2610,
CPT-91/P.2611; for the simplest presentation see
CPT-92/P2814.

\item{[9]} R. Coquereaux, G. Esposito-Far\'ese, G. Vaillant,
{\sl Nucl. Phys.} {\bf B353}  (1991) 689;\br
M. Dubois-Violette, R. Kerner, J. Madore, {\sl J. Math.
Phys.} {\bf 31} (1990) 316;\br
B. Balakrishna, F. G\"ursey and K. C. Wali, {\sl Phys. Lett.}
{\bf 254B} (1991) 430;\br
R. Coquereaux, G. Esposito-Far\'ese and F. Scheck, Preprint
IHES/P/90/100 and CPT-90/PE 2464.

\item{[10]} A.H. Chamseddine, G. Felder and J. Fr\"ohlich
{\sl Phys.Lett.} {\bf296B} (1993) 109, and Zurich preprint
ZU-TH-30/92, to appear in {\sl Nucl. Phys.} {\bf B}.

\item{[11]} M. Gell-Mann, P. Ramond and R. Slansky in
{\sl Supergravity} editor D. Freedman et al (North-Holland,
Amsterdam 1979).

\item{[12]}
 R. Mohapatra and G. Senjanovic {\sl Phys. Rev. Lett.} {\bf
44} (1980) 912;\br
E. Witten {\sl Phys.Lett.} {\bf 91B} (1980) 81.

\item{[13]}A.H. Chamseddine, G. Felder and J. Fr\"ohlich
{\sl Zurich preprint} ETH-TH/92/44, to appear in {\sl Comm.Math.Phys}.

\end